\documentclass{PoS}

\usepackage{graphicx}
\usepackage{amsmath}
\usepackage{amsfonts}
\usepackage{amssymb}
\usepackage{bbm}
\usepackage{xcolor}

\title{Chiral Magnetic Effect in the Dirac-Heisenberg-Wigner formalism}

\ShortTitle{Chiral Magnetic Effect in the Dirac-Heisenberg-Wigner formalism}

\author{D\'aniel Ber\'enyi\\
	Wigner Research Centre for Physics, Hungarian Academy of Sciences, Budapest, Hungary}

\author{\speaker{P\'eter L\'evai}\\
        Wigner Research Centre for Physics, Hungarian Academy of Sciences, Budapest, Hungary\\
        E-mail: \email{levai.peter@wigner.mta.hu}}

\abstract{The emergence of the Chiral Magnetic Effect (CME) and the related anomalous current is investigated using the real time Dirac-Heisenberg-Wigner formalism. This method is widely used for describing strong field physics and QED vacuum tunneling phenomena as well as pair-production in heavy-ion collisions. We extend earlier investigations of the CME in constant flux tube configuration by considering time dependent fields. In our model we can follow the formation of axial charge separation, formation of axial current and then the emergence of the anomalous electric current. Qualitative results are shown for special field configurations that help interpret the predictions of CME related effects in heavy-ion collisions in the RHIC Beam Energy Scan program.}

\FullConference{EPS-HEP 2017, European Physical Society conference on High Energy Physics\\
		5-12 July 2017\\
		Venice, Italy}

\begin{document}

\section{Introduction}

The Quark-Gluon Plasma (QGP) phase is described by Quantum Chromodynamics (QCD) that is a non-Abelian gauge theory. One of the non trivial properties of QCD is that gauge configurations have topological invariants (winding numbers): integer numbers, that are preserved by smooth deformations. In heavy-ion collisions configurations with non-zero winding numbers are expected and these can be interpreted as transitions that may take place as tunneling processes at low temperature (instantons) or as above threshold 'jumps' at high temperature (sphalerons).

The simplest non trivial configuration with a non-zero winding number is the flux tube. When quarks interact with this field they can change their chirality (handedness). Non-central heavy-ion collisions also create very strong magnetic fields, that restrict quarks to the lowest Landau levels, such that they momentum get parallelized/antiparallelized (based on their helicity) with their spin, that is in turn aligned with the magnetic field, and results in a charge separation, that creates a current. So, if due to topological transitions there is an imbalance between chiralities, the separated quarks owing to their charge create an electric current parallel to the magnetic field \cite{KharzeevTopoChargeHIC}.

The process can be formally modeled in the framework of Quantum Electrodynamics (QED) after color diagonalizing the gluon fields, and having the chromoelectric and magnetic fields considered as parallel components of the QED $E$ and $B$ fields while also including a perpendicular $B$ field component for the external magnetic field\cite{Fukushima:2008xe}. We aim to describe the dynamical evolution, so we use the equal-time Wigner-function formalism to calculate the time evolution of the currents in the QED system \cite{BirulaDHW, AlkoferIDHW}. 

\section{The Dirac-Heisenberg-Wigner formalism}

The Wigner-function is a quantum generalization of the classical one-particle phase space density. The Dirac-Heisenberg-Wigner (DHW) formalism gives a relativistic evolution equation, that is only depending on a single time parameter, can be formulated by an initial value problem (starting from vacuum) and suitable to describe the spatio-temporal evolution of a fermionic field under classical external fields.

To model the Chiral Magnetic Effect, the simplest topologically non-trivial configuration is the homogeneous flux-tube, and in that case for static fields it was shown how to decompose the QCD model into a QED analogue problem \cite{FukushimaRealTimeCME}. For time dependent description the QED Wigner function can be used, and can be expanded on the Dirac spinor basis that results in a partial differential equation system of 16 real components. Since we are first interested in describing light quarks, we take the massless limit, and reduce the number of equations to 8. Following the notation of \cite{AlkoferIDHW} the equations read:

\begin{align}
D_t \mathbbm{v}_0&+\vec{D}_{\vec{x}} \cdot \vec{\mathbbm{v}} \phantom{ + 2\vec{p}\times \mathbbm{\vec{a}}}= 0\;, \\
D_t \mathbbm{a}_0&+\vec{D}_{\vec{x}} \cdot \vec{\mathbbm{a}} \phantom{+ 2\vec{p}\times \mathbbm{\vec{a}}}= 0\;, \\
D_t \mathbbm{\vec{v}}&+\vec{D}_{\vec{x}} \mathbbm{v}_0 + 2\vec{p}\times \mathbbm{\vec{a}} = 0\;,\\
D_t \mathbbm{\vec{a}}&+\vec{D}_{\vec{x}} \mathbbm{a}_0 + 2\vec{p}\times \mathbbm{\vec{v}} = 0\;.
\end{align}

where the evolution operators are given without any approximations by $D_t = \partial_t + e\vec{E} \cdot \vec{\nabla}_{\vec{p}}$ and $\vec{D}_{\vec{x}} = e\vec{B} \times \vec{\nabla}_{\vec{p}}$. The components are representing the current density $\mathbbm{\vec{v}}$, the charge density $\mathbbm{v}_0$, the axial current density $\mathbbm{\vec{a}}$ and the axial charge density $\mathbbm{a}_0$.

The initial conditions for vacuum are only non-vanishing for the current density:
\begin{eqnarray}
\vec{\mathbbm{v}}(\vec{p}, t=-\infty) &=& -\frac{2\vec{p}}{\sqrt{m^2+\vec{p}^2}}\;,
\end{eqnarray}

To further simplify the equations we use the Method of Characteristics \cite{AlkoferIDHW}. We integrate the electric field to obtain the vector potential, and use that to shift the momentum variable $\tilde{\vec{p}} = \vec{p} + e \int \vec{E}(t) dt$ to get rid of the $e\vec{E} \cdot \vec{\nabla}_{\vec{p}}$ term in $D_t$. This way only those momentum derivates remain that are multiplied by the magnetic field in $\vec{D}_{\vec{x}}$.

For the numerical solution a global pseudo-spectral collocation solver was developed that utilizes the Graphical Processing Units (GPUs) for the dense tensor operations, enabling a 30x speedup compared to traditional CPU methods. The momentum space functions are expanded on Rational Chebyshev polynomials\cite{BoydRatCheb1, BoydRatCheb2}, and evolved with 4th order explicit Runge-Kutta stepper. The momentum space integrals $\mathbbm{v}^{\mu}(t) = \frac{1}{(2\pi)^3} \int\limits_{-\infty}^{\infty} {\rm d}p^3 \mathbbm{v}^{\mu}(t, \vec{p})$ and $\mathbbm{a}^{\mu}(t) = \frac{1}{(2\pi)^3} \int\limits_{-\infty}^{\infty} {\rm d}p^3 \mathbbm{a}^{\mu}(t, \vec{p})$ are calculated by Clenshaw-Curtis quadrature \cite{BoydInfiniteIntervalQuadrature}. These quantities are the total electric charge and current as well as the total axial charge and current respectively. Electric charge is conserved, so $\mathbbm{v}_0(t) = 0$, but the axial charge develops a non-zero value, since it is related to the chiral imbalance: $\mathbbm{a}_0( t = +\infty) = N_R-N_L$.

The numerical solver was verified on the two important analythic solutions: the time dependent Sauter electric field case \cite{Kruglov} and the stationary magnetic field solution given in \cite{BirulaDHW}.

\subsection{Sauter field configuration}

\begin{figure}[tb]
	\centering
	\includegraphics[width=0.5\linewidth]{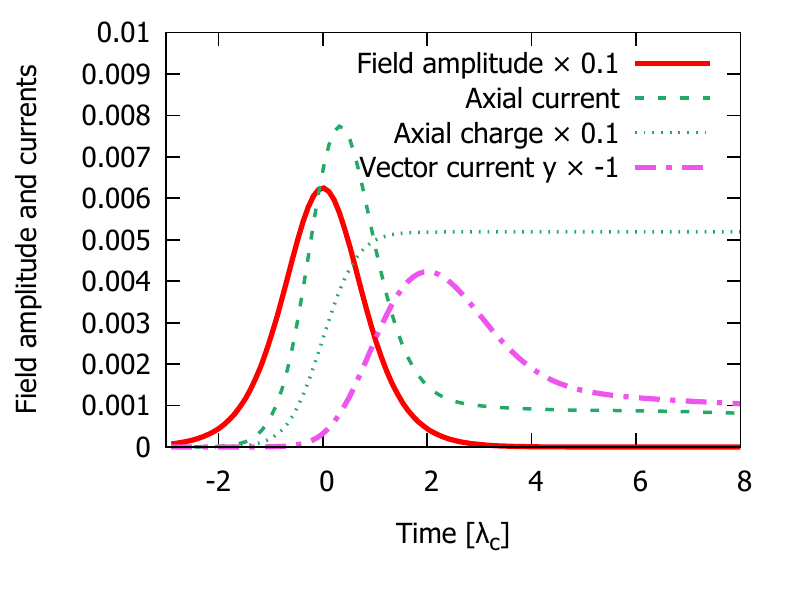}
	\caption{Time dependence of the external field, the axial charge and current and the vector current.}
	\label{fig_th_timedep}
\end{figure}

To verify that this framework can give results consistent with the CME, we first considered a simple time dependent field, the Sauter field $f(t) = A \cosh^{-2}\left(t / \tau\right)$ where the amplitude is measured in critical field units ($E_{cr}$ or $B_{cr}$) and $\tau$ is measured in Compton time $\lambda_c = \frac{\hbar}{mc^2}$ units. We then set the field components to $E_z = B_z = B_y = f(t)$ and all other components to zero. We recorded the momentum space integrals during the time evolution and Figure \ref{fig_th_timedep}. shows the results. Clearly, the chain of events is what is outlined in the introduction: first, the external fields build up. This drives the formation of an axial current, that creates an axial charge separation, that results in the charge displacement that creates the vector current. As the external driving fields decay, the induced charge and currents converge to their asypmtotic values. Note, that even very small external field values are able to sustain the dynamics of the other components. A more detailed investigation was given in \cite{self}.

\subsection{Heavy-Ion Phenomenology}

\begin{figure}[tb]
	\centerline{%
		\includegraphics[width=0.9\linewidth]{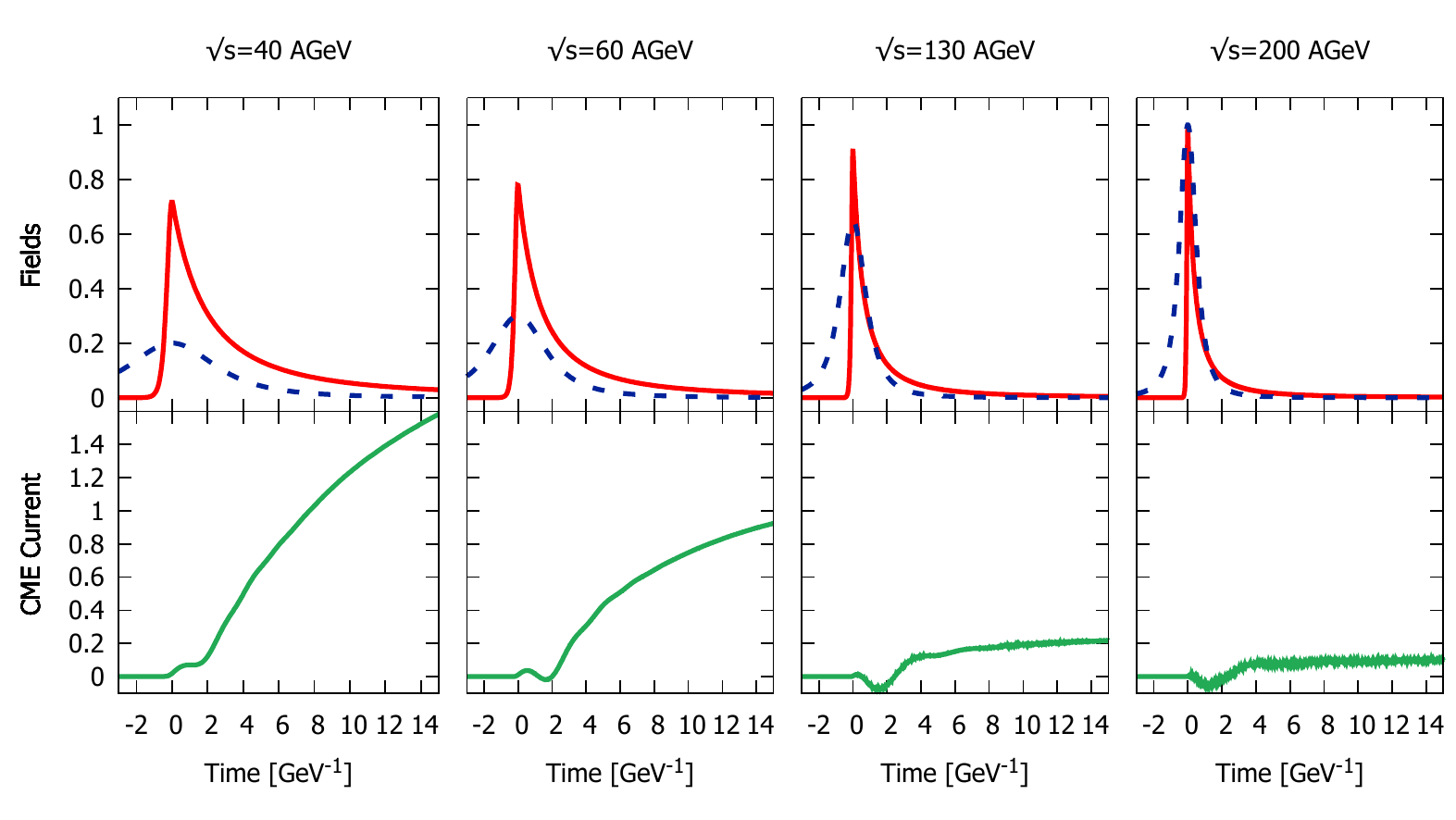}}
	\caption{Time dependence of the external fields (measured in critical field units) and the anomalous CME current at collision energies $\sqrt{s} = 40, 60, 130$ and $200$ AGeV.}
	\label{fig_rhicexp_timedep}
\end{figure}

To model the currents in heavy-ion collision we built a phenomenological model for the external $E$ and $B$ fields. We start with the following time dependent function 

\begin{equation}
\Phi(t, \tau, A, \kappa) = A \cdot
\left\{
\begin{aligned}
&\rm{cosh}^{-2}(10 t/\tau) & & t < 0,\\
&(1+t/\tau)^{-\kappa} & & t \ge 0,
\end{aligned}
\right. \label{eq_Phi}
\end{equation}

where we set $\kappa = 2$ and define the external fields as follows (\cite{SkokovLevaipTspectra} eq. 26. for $E_z, B_z$ and \cite{ZakharovEMRQGP} eq. 12. for $B_y$):

\begin{align}
e\vec{E} &= \{0, 0,                               \qquad\qquad\qquad\quad \Phi(t, \tau, A_{Ez}, \kappa) \}  \;,\\
e\vec{B} &= \{0, A_{By}\left(1+\frac{t^2}{\tau^2}\right)^{-3/2}, \Phi(t, \tau, A_{Bz}, \kappa)  \} \;.
\end{align}


and the other quantities are related to the center of mass energy $\sqrt{s}$ as follows \cite{LevaiSkokovSU2heavyq}:
$\tau =\frac{0.75}{Q_s} \frac{\sqrt{s}_{\rm{RHIC}}}{\sqrt{s}}$, $A_{Ez} = A_{Bz} = Q_s^{2} \left( \frac{\sqrt{s}}{\sqrt{s}_{\rm{RHIC}}} \right)^{\lambda}$, $A_{By} =0.2 Q_s^{2} \frac{\sqrt{s}}{\sqrt{s}_{\rm{RHIC}}}$ with $Q_s = 1 \rm{GeV}$ and $\sqrt{s}_{\rm{RHIC}} = 200 \rm{AGeV}$. The $z$ amplitude scale $\lambda$, is the gluon saturation scale, that is chosen to be 0.2 in accordance with the literature \cite{Golec_Sat, Kowalski_Sat}.

Figure \ref{fig_rhicexp_timedep}. shows the shape of the external fields, and the y component of the vector current for different collision energies. Smaller energies result in larger $\tau$, which is known in the strong field picture to increase the magnitude of currents and make the temporal dynamics last longer. This is clearly observed in the time dependence anomalous electric current, $V_y$, which slowly approaches its asymptotic value, since the driving fields have only a cut-power law decay (c.f. eq. \ref{eq_Phi}).

The unexpected phenomena at almost all energies is that the anomalous current starts with a negative sign and undergoes a reversal at $t \approx 2 \ {\rm GeV^{-1}}$
after the collision. Between 40-60 AGeV the negative dip disappears that we attribute to the larger width of the pulse. Precise values depend on the external field models, it's amplitude, gradients, decay rate, but for pulse like fields the overall behavior is the same.

\section{Discussion}

Our calculation based on the real-time Dirac-Heisenberg-Wigner formalism
has shown that the RHIC Beam Energy Scan program should be capable 
to observe the CME effect with changing the bombarding energy. Unfortunately
our results indicate the disappearance of the effect at the highest RHIC energies,
as well as at LHC energies.

\end{document}